\documentclass[preprint]{aastex}
\begin{document}
\title{Distances and lensing in cosmological void models}
\author{Kenji Tomita}
\affil{Yukawa Institute for Theoretical Physics, Kyoto University,
Kyoto 606-8502, Japan}
\email{tomita@yukawa.kyoto-u.ac.jp}
\begin{abstract}
We study the distances and gravitational lensing in spherically 
symmetric inhomogeneous cosmological models consisting of inner and
outer homogeneous regions which are connected by a single shell 
or double shells at the redshift $z_1 \sim 0.067$. The density and
Hubble parameters in the inner region are assumed to be smaller and
larger, respectively, than those in the outer region. It is found that 
at the stage $z_1 < z < 1.5$ the distances from an observer in the 
inner void-like region are larger than the counterparts (with equal 
$z$) in the corresponding homogeneous Friedmann models, and 
hence the magnitudes for the sources at this stage are larger. 
This effect of the void-like low-density region may explain the deviations of 
the observed [magnitude-redshift] relation of SNIa from the relation 
in homogeneous models, independently of the cosmological constant. 
When the position of the observer deviates from the center, moreover, 
it is shown that the distances are anisotropic and the images of remote 
sources are systematically deformed. The above relation at $z \gtrsim
1.0$ and this anisotropy will observationally distinguish the role 
of the above void-like region from that of the positive cosmological 
constant. The influence on the time-delay measurement is also discussed.
\end{abstract}

\keywords{cosmology: large scale structure of the universe - observations}

\section{Introduction}
\label{sec:level1}

In recent cosmological observations, the following three remarkable
phenomena have been discovered, which may contradict with the
homogeneity of the universe. One of them is the large-scale bulk flows 
in the region with distance $< 150 h^{-1}$ Mpc \ $(H_0 = 100 h^{-1}$
km sec$^{-1}$ Mpc$^{-1})$ without the associated large CMB dipole
anisotropy  \citep{hud99, will99}.
 Second we have the inhomogeneity of the observed Hubble 
constant whose values are smaller in the measurements for remoter
sources \citep{bran98, free97, san97, blan97}.
 The last one is the magnitude-redshift relation of SNIa, in
which the observed magnitudes of sources are larger than those
expected in the homogeneous Friedmann models without the cosmological
constant. For the model-fitting the positive cosmological constant
which brings an ``accelerating'' universe was considered as a
necessary quantity \citep{riess98,schmidt98,garna98}.

For the explanation of the first phenomenon we considered spherically
symmetric inhomogeneous models in a previous paper \citep{tom99},
which is cited as paper I in the following. They consist of inner 
and outer homogeneous regions connected with a single shell, double 
shells or an intermediate self-similar region (Tomita 1996, 1995), 
and it is assumed that the density
and Hubble parameters in the inner region are smaller and larger,
respectively, than those in the outer region. Then it was shown that
the observed situation of bulk flows and CMB dipole anisotropy can be
reproduced, if the radius of the boundary of the two region and the 
observer's position from the center are about $200 h^{-1}$ Mpc and 
$40 h^{-1}$ Mpc, respectively. The consistency with the observed bulk
flows \citep{hud99, will99, dale99, giov98} was shown in paper I. 
These models are found to be consistent also with the observed 
inhomogeneity of the Hubble constant. 

The inhomogeneity of the Hubble constant has already been discussed by 
various workers (e.g. \citet{turn92}, \citet{bart95}). 
The local void region with higher Hubble constant was studied
independently by \citet{zehav98} as
the local Hubble bubble, which has the scale $\sim 70 h^{-1}$ Mpc and
is bordered by the dense walls as the Great Attractor. 
They analyzed the statistical relation 
between the distances and the local Hubble constants derived from the
data of SNIa, and found the existence of a void region with a local
Hubble constant larger than the global Hubble constant.
The relation to the SNIa data on the scale $\sim 150 h^{-1}$ Mpc will 
be discussed in this paper from our standpoint.

In the present paper the behavior of the distances is investigated in 
the models in the previous paper with similar model parameters. In \S
2, we treat the
distances from a virtual observer who is in the center of the inner
void-like region in models with a single shell, and derive the 
[magnitude $m$ - redshift $z$] relation. This relation is compared
with the counterpart in the homogeneous models. Then the relation 
in the present models is found to deviate from that in the homogeneous
models with $\Lambda = 0$ at the stage of $z < 1.5$. It is partially 
similar to that in the nonzero-$\Lambda$ homogeneous models,
but the remarkable difference appears at the high-redshift stage $z > 1.0$. 
In \S 3, we consider a realistic observer who is in the
position deviated from the center, and calculate the distances from
him. The distances depend on the direction of incident light and the
area angular diameter distance is different from the linear angular diameter
distances. It is shown as the result that the [$m, z$] relation is
anisotropic, but the relation averaged with respect to the angle is
very near to the relation by the virtual observer.
The comparison with the observed relation for SNIa is also
discussed. In \S 4 we derive the lens effect such as the convergence 
and shear of the images caused by the above anisotropic nature of
distances, and in \S 5 discuss the influence on the time-delay from 
a remote double quasar. \S 6 is dedicated to concluding remarks.
In Appendix A, the derivation of distances in models with double
shells is described in parallel with \S 2 and \S 3.

\section{Distances from the center of the inner region}
\label{sec:level2}

In this section we treat the models with a single shell, in which 
the inner
homogeneous region V$^{\rm I}$ and the outer homogeneous region
V$^{\rm II}$ are connected with a shell and the lineelement is
expressed as 
\begin{equation}
  \label{eq:m1}
 ds^2 = g^j_{\mu\nu} (dx^j)^\mu (dx^j)^\nu \\ = - c^2 (dt^j)^2 + [a^j
 (t^j)]^2 \Big\{ d (\chi^j)^2 + [f^j (\chi^j)]^2 d\Omega^2 \Big\},
\end{equation}
where $j \ (=$ I or II) represents the regions, $f^j (\chi^j) = \sin
\chi^j, \chi^j$ and $\sinh \chi^j$ for $k^j = 1, 0, -1$, respectively,
and $d\Omega^2 = d\theta^2 + \sin^2 \theta \varphi^2$. 
The model parameters are expressed as $H_0^j \ (= 100 h_0^j),
 \Omega_0^j$ and $\lambda_0^j$.
Here the negative curvature is assumed in all regions. 
In the paper I, we showed the Einstein equations in both regions and
the junction conditions at the boundary shell which is given as 
$\chi^{\rm I} = \chi^{\rm I}_1$ and $\chi^{\rm II} = \chi^{\rm II}_1$, 
and derived the equations of light rays in both regions. 

Using the latter equations we obtained CMB dipole and quadrupole
anisotropies and found that the influence of motions of the shell on
the light paths is negligibly small, so that the approximation of the
comoving shell is good for the treatment of light rays. In  this paper 
this approximation is assumed throughout in all cases, and the
equations in the paper I are cited as Eq. (I.1), Eq.(I.A1) and so on.

The angular diameter distance $d_A$ between a virtual observer at the
center O (in the inner region V$^{\rm I}$) and the sources S is given
by 
\begin{equation}
  \label{eq:m2}
d_A = a^{\rm I}(\bar{\eta}^{\rm I}_{\rm s}) \sinh (\bar{\chi}^{\rm I}_{\rm
s}), 
\end{equation}
if S is in V$^{\rm I}$, where $(\bar{\eta}^{\rm I}_{\rm s},
\bar{\chi}^{\rm I}_{\rm s})$ are the coordinates of S. Here and in the 
following, bars are used for the coordinates along the light paths to
the virtual observer, as in the paper I. If S is in V$^{\rm II}$, we
have
\begin{equation}
  \label{eq:m3}
d_A = a^{\rm I}(\bar{\eta}^{\rm I}_1) \sinh (\bar{\chi}^{\rm I}_1) 
+ [a^{\rm II}(\bar{\eta}^{\rm II}_{\rm s} \sinh (\bar{\chi}^{\rm II}_{\rm
s}) - a^{\rm II}(\bar{\eta}^{\rm II}_1) \sinh (\bar{\chi}^{\rm II}_1)] 
= a^{\rm II}(\bar{\eta}^{\rm II}_{\rm s}) \sinh (\bar{\chi}^{\rm II}_{\rm
s}), 
\end{equation}
where Eq. (I.2) for $R$ was used. When S is in V$^{\rm I}$, 
$(\bar{\eta}^{\rm I}_{\rm s}, \bar{\chi}^{\rm I}_{\rm s})$ are related 
to $(\bar{\eta}^{\rm I}_0, 0)$ by
\begin{equation}
  \label{eq:m4}
\bar{\eta}^{\rm I}_0 - \bar{\eta}^{\rm I}_{\rm s} = \bar{\chi}^{\rm I}_{\rm
s}, 
\end{equation}
where $\bar{\eta}^{\rm I}_0$ is given by Eqs. (I.13) and (I.14)
with $y^{\rm I} = 1$, and $\bar{\eta}^{\rm I}_{\rm s}$ is related to
the source redshift $\bar{z}^{\rm I}_{\rm s}$ by
\begin{equation}
  \label{eq:m5}
1 + \bar{z}^{\rm I}_{\rm s} = 2(1-\Omega_0^{\rm I})/\Omega_0^{\rm I}
/(\cosh \bar{\eta}^{\rm I}_{\rm s} -1).
\end{equation}
For a given $\bar{z}^{\rm I}_{\rm s}$, we obtain $\bar{\eta}^{\rm I}_{\rm s}$
and $\bar{\chi}^{\rm I}_{\rm s}$, and hence $d_A$ from
Eq. (\ref{eq:m2}). 

When S is in V$^{\rm II}$, we have
\begin{equation}
  \label{eq:m6}
\bar{\eta}^{\rm I}_0 - \bar{\eta}^{\rm I}_1 = \bar{\chi}^{\rm I}_1 
\end{equation}
and
\begin{equation}
  \label{eq:m7}
\bar{\eta}^{\rm II}_1 - \bar{\eta}^{\rm I}_{\rm s} = \bar{\chi}^{\rm
II}_{\rm s} - \bar{\chi}^{\rm II}_1, 
\end{equation}
where $\bar{\eta}^{\rm I}_1$ is related to $\bar{z}^{\rm I}_1$ by
\begin{equation}
  \label{eq:m8}
1 + \bar{z}^{\rm I}_1 = 2(1-\Omega_0^{\rm I})/\Omega_0^{\rm I}
/(\cosh \bar{\eta}^{\rm I}_1 -1)
\end{equation}
and so $\bar{\chi}^{\rm I}_1$ is also related to $\bar{z}^{\rm I}_1$
using Eq. (\ref{eq:m6}). Coordinates at the shell , $(\bar{\eta}^{\rm
I}_1, \bar{\chi}^{\rm I}_1)$ and $(\bar{\eta}^{\rm II}_1,
\bar{\chi}^{\rm II}_1)$ are connected by Eqs. (I.53) and
(I.54). On the other hand, $\bar{\eta}^{\rm II}_{\rm s}$ is related
to $\bar{z}^{\rm II}_{\rm s}$ by
\begin{equation}
  \label{eq:m9}
{1 + \bar{z}^{\rm II}_{\rm s} \over 1 + \bar{z}^{\rm II}_1} = 
{\cosh \bar{\eta}^{\rm II}_1 -1 \over \cosh \bar{\eta}^{\rm II}_{\rm s} - 1},
\end{equation}
where $\bar{z}^{\rm II}_1$ is equal to $\bar{z}^{\rm I}_1$ under the
approximation of the comoving shell (cf. Eq. (I.6)). Accordingly,
$d_A$ is uniquely determined for given $\bar{z}^{\rm I}_1 \ (=
\bar{z}^{\rm II}_1)$ and $\bar{z}^{\rm II}_{\rm s}$. 

The distance ($d^F_A$) in a homogeneous Friedmann model is 
\begin{equation}
  \label{eq:m10}
d^F_A = a^{\rm I} (\bar{\eta}^{\rm I}_{\rm s}) \sinh \bar{\chi}^{\rm
I}_{\rm s}, 
\end{equation}
which is defined in V$^{\rm I}$ for arbitrary $\bar{z}^{\rm I}_1$,
assuming that V$^{\rm I}$  covers every region of the model.

The luminosity distances $d_L$ and $d_L^F$ are defined in terms of $d_A$ 
$d_A^F$ as $d_L = (1 + z)^2 d_A$ and $d_L^F = (1 + z)^2 d_A^F.$ 
Accordingly the ratio of luminosity 
distances is equal to the ratio of angular diameter distances.
In Figs. 2 and 3 the behavior of $5 \log d_L$ as a function of
$z \ (= \bar{z}^j_{\rm s})$ is shown in the case of $\bar{z}^{\rm I}_1 =
0.067$ for the parameters $(\Omega^{\rm I}_0, \Omega^{\rm II}_0, h^{\rm
 I}, h^{\rm II}/h^{\rm I}) = (0.2, 0.56, 0.7, 0.82)$ and $(0.2, 0.88, 
0.7, 0.82)$, respectively, which are appropriate to describe the bulk
flow in the previous paper \citep{tom99}. The lines in the homogeneous 
models with ($\Omega_0, \lambda_0) = (0.2, 0)$ and $(0.2, 0.8)$ are 
also shown for
comparison. It is found that $5 \log d_L$ is larger than that in the
model with $(0.2, 0)$ for $\bar{z}_1^{\rm I} < z < 1.5$, it is
consistent with that in the model with $(0.2, 0.8)$ for $0.1 < z <
0.3$, and it is intermediate for $0.1 < z < 1.0$ between the two
models with $(0.2, 0)$ and $(0.2, 0.8)$. The difference between 
the present shell model and the non-zero $\lambda_0$ model is 
remarkable for $z > 1.0$. 
According to recent data of high-redshift SNIa \citep{riess98,
schmidt98,garna98},  it is at epochs of $z \sim 0.4$ that 
their data of $m$ deviate conspicuously from the values in the
homogeneous models with vanishing cosmological constant. From the
comparison with their data, it seems that the present models can 
explain their data as well as the non-zero $\lambda_0$ model.

In Fig. 4 the behavior of $5 \log d_L$ (by the observer at C) in 
the double-shell models is shown for
$\bar{z}_1^{\rm I} = 0.05$ and $\bar{z}_2^{\rm II} = 0.1$, where the
distances in the double-shell case are derived in the Appendix A.
It is found that the general trend is same as that in the single-shell case.  

In Figs. 5 and 6, the magnitude differences $\Delta m \ [\equiv 5 \log
d_L - 5 \log (d_L)_{\rm homog}]$ are shown in the single-shell and 
double-shell cases, respectively. In Fig. \ref{fig5} the difference 
for $\Omega_0 = 0.3$ is also shown and it is found to be smaller than 
that for $\Omega_0 = 0.2$.

\section{Distances from a non-central observer O in the inner region}
\label{sec:level3}

Next let us derive the angular diameter distance $d_A$ from an
observer O whose position is deviated from the center. When the source 
S is in V$^{\rm I}$, \ $d_A$ is equal to $d_A^F$ in the homogeneous
Friedmann model. In the following we consider the case when S is in
the outer regions. In this case we have two linear angular diameter
distances (the longitudinal linear distance $d_A^l$ and the transverse 
linear distance $d_A^t$) for the angles in the $\varphi$ and $\theta$
directions, respectively, and the area angular diameter distance
$d_A^a \ [\equiv (d_A^l d_A^t)^{1/2}]$ \ (cf. Fig. \ref{fig7}).

Let us consider a single-shell model and assume that, in the plane 
$\theta =$ const, two light rays
start from S at $(\eta^{\rm II}_{\rm s}, \chi^{\rm II}_{\rm s},
\varphi_{\rm s})$ and $(\eta^{\rm II}_{\rm s}, \chi^{\rm II}_{\rm s},
\varphi_{\rm s} + \Delta \varphi_{\rm s})$  and reach O at 
$(\eta^{\rm I}_0, \chi^{\rm I}_0, 0)$ with the angle $\phi$ and $\phi
+ \Delta \phi$, respectively. If the angular diameter distance from S
to the center C is $d_A(\eta^{\rm II}_{\rm s}, \chi^{\rm II}_{\rm
s})$, the proper interval of the two rays in S is equal to
$d_A(\eta^{\rm II}_{\rm s}, \chi^{\rm II}_{\rm s}) d \varphi_{\rm s}$, 
so that the longitudinal linear distance is defined by
\begin{equation}
  \label{eq:r1}
d_A^l \equiv d_A(\eta^{\rm II}_{\rm s}, \chi^{\rm II}_{\rm s}) \
\partial  \varphi_{\rm s}/ \partial \phi /[\cos (\phi -
\varphi_{\rm s})],
\end{equation}
where $d_A(\eta^{\rm II}_{\rm s}, \chi^{\rm II}_{\rm s})$ is given by
\begin{equation}
  \label{eq:r2}
d_A(\eta^{\rm II}_{\rm s}, \chi^{\rm II}_{\rm s}) = a^{\rm
II}(\eta^{\rm II}_{\rm s}) \sinh (\chi^{\rm II}_{\rm s}).
\end{equation}

Next let us consider two rays with equal $\phi$ in planes of $\theta = 
0$ and $\Delta \theta$, when $\Delta \theta << \pi$. Then [the angle 
between two rays reaching O] \ 
(= $\Delta \theta \sin \varphi_{\rm s}$) is equal to [the angle
between two rays reaching C] multiplied by a factor ($\sin 
\phi/\sin \varphi_{\rm s}$). Therefore the transverse linear distance
is
\begin{equation}
  \label{eq:r3}
d_A^t = d_A(\eta^{\rm II}_{\rm s}, \chi^{\rm II}_{\rm s}) \sin
\varphi_{\rm s}/ \sin \phi.
\end{equation}
In the following we derive the relations between 
$(\eta^{\rm II}_{\rm s}, \chi^{\rm II}_{\rm s},
\varphi_{\rm s}), \ (\eta^{\rm I}_0, \chi^{\rm I}_0, 0)$ and the
incident angle $\phi$. When we give the redshift $\bar{z}_1^{\rm I}$, the
radial coordinate $\chi^{\rm I}_1$ is fixed using Eqs. (\ref{eq:m6})
and (\ref{eq:m8}).

In V$^{\rm I}$ we have Eq. (I.40) for $\eta^{\rm I}_1$ in the case
of $\phi = \phi_1$ and $\pi - \phi_1$, and 
\begin{equation}
  \label{eq:r4}
h_0^{\rm I} = [1 + (\zeta^{\rm I})^2]^{1/2}, \quad \zeta^{\rm I} =
\sinh \chi^{\rm I}_0 \sin \phi,
\end{equation}
where $\chi^{\rm I}_0$ is fixed by giving the distance CD, that is,
\begin{equation}
  \label{eq:r5}
l_0 = a_0 \chi_0^{\rm I}.
\end{equation}
For $(\eta^{\rm II}_{\rm s}, \chi^{\rm II}_{\rm s})$ we obtain from
Eq. (I.41)
\begin{equation}
  \label{eq:r6}
G(\chi^{\rm II}_{\rm s}) \equiv \cosh^{-1} \Big({\cosh
\chi^{\rm II}_{\rm s} \over
h_0^{\rm II}}\Big) - \cosh^{-1} \Big({\cosh \chi_1^{\rm II} \over
h_0^{\rm II}}\Big) = \eta_1^{\rm II} - \eta_{\rm s}^{\rm II}.
\end{equation}
 The definition (I.37) of $h_0^j$ and the junction conditions
 (I.46) and (I.47) lead to 
\begin{equation}
  \label{eq:r7}
h_0^{\rm II} = [1 + (\zeta^{\rm II})^2]^{1/2}, \quad \zeta^{\rm II} =
{a_0^{\rm I} \over a_0^{\rm II}} \zeta^{\rm I},
\end{equation}
where $A_0^j \ (j =$ I, II) are given by Eq. (I.12). The coordinates 
at the boundary $(\eta^{\rm II}_1, \chi^{\rm II}_1)$ in
Eq. (\ref{eq:r6}) are related to $(\eta^{\rm I}_1, \chi^{\rm I}_1)$,
using 
\begin{equation}
  \label{eq:r8}
a_0^{\rm I} y^{\rm I} ({\eta}_1^{\rm I}) \sinh \chi_1^{\rm I} = a_0^{\rm
II} y^{\rm II} ({\eta}_1^{\rm II}) \sinh \chi_1^{\rm II} 
\end{equation}
and 
\begin{equation}
  \label{eq:r9}
a_0^{\rm I} \int_0^{{\eta}_1^{\rm I}} y^{\rm I} (\eta) 
d\eta  = a_0^{\rm II} \int_0^{{\eta}_1^{\rm II}} y^{\rm II} 
(\eta) d\eta,
\end{equation}
which are derived from Eq. (I.2) for $R$ and Eq. (I.6) with
$\gamma^{\rm I} = \gamma^{\rm II} = 1$.  Therefore, $\eta_{\rm s}^{\rm
II}, \chi_{\rm s}^{\rm II}$ are determined by specifying the values of 
$\bar{z}_1^{\rm I}, \ \bar{z}_{\rm s}^{\rm II}$ and $\phi_1$.

Next let us derive $\varphi$ by integrating the ray equations 
\begin{equation}
  \label{eq:r10}
{d \varphi \over d \chi^j} = {(k^\varphi)^j \over (k^\chi)^j},
\end{equation}
where $(k^\varphi)^j$ and $(k^\chi)^j$ for $j =$ I and II are shown
in Eqs. (I.32) and (I.33). The solution of (\ref{eq:r10}) satisfying 
the conditions that $\varphi (\chi = \chi_0) = 0$ and $\varphi (\chi > 0)
\rightarrow \phi$ for $\chi_0 \rightarrow 0$ \ is 
\begin{eqnarray}
  \label{eq:r11}
 \varphi = &&\tan^{-1} \Big\{{1 \over \zeta^{\rm I}} 
\Big[\sinh^2 \chi^{\rm I} + \cosh \chi^{\rm I} \sqrt{\sinh^2 \chi^{\rm
I} - (\zeta^{\rm I})^2}\Big]\Big\}\cr
&& - \tan^{-1} \Big\{{1 \over \zeta^{\rm I}} 
\Big[\sinh^2 \chi_0^{\rm I} + \cosh \chi_0^{\rm I} \sqrt{\sinh^2 
\chi_0^{\rm I} - (\zeta^{\rm I})^2}\Big]\Big\}
\end{eqnarray}
and 
\begin{eqnarray}
  \label{eq:r12}
 \varphi = &&\tan^{-1} \Big\{{1 \over \zeta^{\rm I}} 
\Big[\sinh^2 \chi^{\rm I} - \cosh \chi^{\rm I} \sqrt{\sinh^2 \chi^{\rm
I} - (\zeta^{\rm I})^2}\Big]\Big\}\cr 
&& - \tan^{-1} \Big\{{1 \over \zeta^{\rm I}} 
\Big[\sinh^2 \chi_0^{\rm I} - \cosh \chi_0^{\rm I} \sqrt{\sinh^2 
\chi_0^{\rm I} - (\zeta^{\rm I})^2}\Big]\Big\}.
\end{eqnarray}
Solutions (\ref{eq:r11}) and (\ref{eq:r12}) are applicable for $k^j >
0$ and $< 0$, respectively.

In V$^{\rm I}$ we have for $\phi = \phi_1$
\begin{equation}
  \label{eq:r13}
\varphi =  \tan^{-1} \Big\{{1 \over \zeta^{\rm I}} 
\Big[\sinh^2 \chi^{\rm I} + \cosh \chi^{\rm I} \sqrt{\sinh^2 \chi^{\rm
I} - (\zeta^{\rm I})^2}\Big]\Big\},
\end{equation}
\begin{equation}
  \label{eq:r14}
\varphi_1 = \varphi (\chi^{\rm I} = \chi_1^{\rm I}).
\end{equation}
For $\phi = \pi - \phi_1$
\begin{eqnarray}
  \label{eq:r15}
\varphi = &&\tan^{-1} \Big\{{1 \over \zeta^{\rm I}} 
\Big[\sinh^2 \chi^{\rm I} - \cosh \chi^{\rm I} \sqrt{\sinh^2 \chi^{\rm
I} - (\zeta^{\rm I})^2}\Big]\Big\}\cr
&& - \tan^{-1} \Big\{{1 \over \zeta^{\rm I}} 
\Big[\sinh^2 \chi_0^{\rm I} - \cosh \chi_0^{\rm I} \sqrt{\sinh^2 
\chi_0^{\rm I} - (\zeta^{\rm I})^2}\Big]\Big\},
\end{eqnarray}
\begin{equation}
  \label{eq:r16}
\varphi_m = \varphi (\chi^{\rm I} = hi^{\rm I}_m)  
\end{equation}
in the interval $\chi_0 < \chi \leq \chi_m$, and
\begin{eqnarray}
  \label{eq:r17}
\varphi = \varphi_m &+& \tan^{-1} \Big\{{1 \over \zeta^{\rm I}} 
\Big[\sinh^2 \chi^{\rm I} + \cosh \chi^{\rm I} \sqrt{\sinh^2 \chi^{\rm
I} - (\zeta^{\rm I})^2}\Big]\Big\} \cr
&-& \tan^{-1} \Big\{{1 \over \zeta^{\rm I}} 
\Big[\sinh^2 \chi^{\rm I}_m + \cosh \chi^{\rm I}_m \sqrt{\sinh^2 \chi^{\rm
I}_m - (\zeta^{\rm I})^2}\Big]\Big\},
\end{eqnarray}
\begin{equation}
  \label{eq:r18}
\varphi_1 = \varphi (\chi^{\rm I} = \chi^{\rm I}_1)
\end{equation}
in the interval $\chi_m < \chi \leq \chi_1$.

In V$^{\rm II}$ we have  
\begin{eqnarray}
  \label{eq:r19}
\varphi = \varphi_1 &+& \tan^{-1} \Big\{{1 \over \zeta^{\rm II}} 
\Big[\sinh^2 \chi^{\rm II} + \cosh \chi^{\rm II} \sqrt{\sinh^2 \chi^{\rm
II} - (\zeta^{\rm II})^2}\Big]\Big\} \cr
&-& \tan^{-1} \Big\{{1 \over
\zeta^{\rm II}} \Big[\sinh^2 \chi^{\rm II}_1 + \cosh \chi^{\rm II}_1 
\sqrt{\sinh^2 \chi^{\rm II}_1 - (\zeta^{\rm II})^2}\Big]\Big\},
\end{eqnarray}
\begin{equation}
  \label{eq:r20}
\varphi_{\rm s} = \varphi (\chi^{\rm II} = \chi^{\rm II}_{\rm s}).
\end{equation}
Thus $\varphi_{\rm s}$ was derived as a function of $\bar{z}_1^{\rm
I}, \ z_{\rm s}^{\rm II}$ and $\phi \ (= \phi_1$ or $\pi -  \phi_1)$.
$d_A^l, \ d_A^t$ and $d_A^a$ depend on the angle $\phi$. 
Here we calculate the average value of $d_A^a$ defined by
\begin{equation}
  \label{eq:r21}
\langle d_A^a \rangle = {1 \over 2} \int^\pi_0 d_A^a \sin \phi d \phi.
\end{equation}
The $z$ dependence of average values of the corresponding luminosity
distance $d_L^a \ (= (1+z)^2 d_A^a)$ is shown in Figs. \ref{fig2} and 
\ref{fig3} in  comparison with the distance by the
observer C. It is found that, in almost all range of $z$, two lines
indicating these two distances are overlapped and cannot be
distinguished. As for the average behavior of distances from the observer
O, therefore, we can approximately use the distances from the observer C 
for the former ones.

\section{Lensing due to the inhomogeneity}

When the observer's position (O) deviates from the center (C) of the inner
region, the area distance from the observer is anisotropic and the two 
linear distances are not equal. This is a lens effect caused by the
anisotropic and inhomogeneous matter distribution around the
observer (O). Here we discuss the $\phi$ dependence  of the area distance 
and the ratio of the two linear distances $d_A^l/d_A^t$.

\subsection{Area angular diameter distance}

The angular diameter distance is largest and smallest in the 
directions of $\phi = 0$ and $\pi$, respectively. This is because 
light in the directions of $\phi = 0$ and $\pi$ spends the longest and 
shortest time in V$^{\rm II}$, respectively. The behavior of $m$ is 
also similar. Here we consider
$m$ for the distance and treat the magnitude difference $\Delta m \ 
(\equiv m - m_{\rm homog})$, where $m_{\rm homog}$ is the magnitude in 
the Friedmann model with the same $\Omega_0$.
The $z$ dependence of the magnitude difference $\Delta m$ 
for the observer O was derived in the single-shell models.
In Figs. \ref{fig8} and \ref{fig9}, the values averaged for $\phi < 
\pi/4, \pi/4 < \phi < 3\pi/4,$ and $\phi > 3\pi/4$ are shown. The
difference between $\Delta m (\phi < \pi/4)$ and $\Delta m (\phi >
3\pi/4)$ reaches $\sim 0.4$ mag  at the epoch $z \sim 0.1$. 
This difference may represent large dispersions in $m$ around this epoch.
To confirm observationally whether there exists this 
anisotropy in $\Delta m$ actually is important to clarify the validity 
of the present models.

\subsection{Ellipticity of image deformation}

Since $d_A^l$ and $d_A^t$ are different for the sources of $z >
z_1^{\rm I}$, their images are deformed and the degrees of deformation 
depend on $\phi$. Here we define the ellipticity $e$ by
\begin{equation}
  \label{eq:len2}
{1 + e \over 1 - e} = {d_A^l \over d_A^t} \quad {\rm or} \quad  
e = {d_A^l - d_A^t \over d_A^l + d_A^t}. 
\end{equation}
The $z$ dependence of $e$ was calculated in two model parameters for
the study of its general behavior. The maximum and minimum of $e$ are in
$\phi\sim \pi/2$ and $\phi = 0, \pi$, respectively. To clarify the $\phi$
dependence of $e$, we derived the average value for $ 0 < \phi < \pi$, and 
the values averaged for $\phi < \pi/4, \pi/4 < \phi < 3\pi/4,$ and 
$\phi > 3\pi/4$. Their results are shown in
Figs. \ref{fig10} and \ref{fig11}. It is found that the ellipticity
$e$ increases abruptly directly after the epoch $z = z_1^{\rm I}$ and
decreases gradually with $z$.   Accordingly, we should measure the
images of the sources around $z = z_1^{\rm I}$ to confirm the lens effect.

\section{Time-delay for a remote double quasar}

The time-delay for a remote double source is basically caused by the
geometrical length difference and gravitational redshift around the
lens object, and so the formula in the present situation is the same as
that in the homogeneous models. It is expressed \citep{sas93, bland86, 
schn92} as
\begin{equation}
  \label{eq:t1}
\Delta t = \Delta t_{\rm geom} + \Delta t_{\rm grav},
\end{equation}
where
\begin{equation}
  \label{eq:t2}
\Delta t_{\rm geom} = {\cal D} C_{\rm geom}, \ \ {\cal D} \equiv {D_{\rm l}
D_{\rm s} \over D_{\rm ls}}, \ \ C_{\rm geom} \equiv {1 \over 2} (1+z_{\rm l}) 
(\mbox{\boldmath $\theta$} - \mbox{\boldmath $\phi$})^2,  
\end{equation}
\begin{equation}
  \label{eq:t3}
\Delta t_{\rm grav} = (D_{\rm l})^2 C_{\rm grav}, \ \ C_{\rm grav} \equiv - 
{1 \over \pi} (1+z_{\rm l}) \int d \mbox{\boldmath $\theta'$}^2 \Sigma
( \mbox{\boldmath $\theta'$}) \ln [|\mbox{\boldmath $\theta$} - 
\mbox{\boldmath $\theta'$}|/\theta_c].
\end{equation}
Angular diameter distances $D_{\rm l}, D_{\rm s}$ and $D_{\rm ls}$ denote the
distances between an observer O and a lens L, 
between O and a source S, and between L and S, respectively. 
The angle vector $\mbox{\boldmath $\theta$}$ indicates the
position of the ray relative to L in the lens plane, $ \theta_c$
denotes the critical angle of the lens object, and $\Sigma (\mbox{\boldmath
$\theta$})$ is the surface density. Another angle vector
$\mbox{\boldmath $\phi$}$
indicates the position of S relative to L.  For a double quasar we
have the difference of time-delays $\delta \Delta t \equiv {\cal D} \delta 
C_{\rm geom} + (D_{\rm l})^2 \delta C_{\rm grav}$, where the differences of
coefficients $\delta C_{\rm geom}$ and $\delta C_{\rm grav}$ are
determined if the relative positions of L and S and the mass
distribution in L are given.

If $z_s > z_1^{\rm I}$, both L and S are in V$^{\rm II}$, so that
in the above formulas $D_{\rm ls}$ is given by $D_{\rm ls} = D_A 
(\Omega_0^{\rm II}, H_0^{\rm II}, z_{\rm l}, z_{\rm s})$, as in 
the homogeneous models, where
\begin{eqnarray}
  \label{eq:t5}  
D_A (\Omega_0, H_0, z_{\rm l}, z_{\rm s}) \equiv  {2(c/H_0) \over \Omega_0^2
(1+z_{\rm l})(1+z_{\rm s})^2} && [(2-\Omega_0+\Omega_0 z_{\rm s}) 
(1+\Omega_0 z_{\rm l})^{1/2}\cr
&& - (2-\Omega_0+\Omega_0 z_{\rm l}) (1+\Omega_0 z_{\rm s})^{1/2}].
\end{eqnarray}

If we derive the effective Hubble constant $(H_0)_{\rm eff}$ from the
time-delay measurement, we have the relation $(H_0)_{\rm eff} \propto
1 / \delta \Delta t$. For a homogeneous model with $\Omega_0^{\rm I}$
and $H_0^{\rm I}$, moreover, we have 
the relation $H_0^{\rm I} \propto 1 / [\delta \Delta t]_{\rm homog}$.
From these two relations we obtain  
\begin{equation}
  \label{eq:t4}  
{H_0^{\rm I} \over (H_0)_{\rm eff}} = {\delta \Delta t \over [\delta
\Delta t]_{\rm homog}} = {\alpha_1 + \alpha_2 \beta \over 1 + \beta}, 
\end{equation}
where $\alpha_1 \equiv {\cal D}/{\cal D}_{\rm homog}, \ \alpha_2 \equiv
(D_{\rm l})^2/ (D_{\rm l})^2_{\rm homog},$ and $\beta \equiv [{\cal
D}/(D_{\rm l})^2]_{\rm homog} \delta C_{\rm geom}/\delta C_{\rm grav} \ (> 0)$.
By solving this equation, therefore, we can obtain $H_0^{\rm II}$ 
in the present single-shell models. Since the lensing is caused in 
V$^{\rm II}$, it is natural that we can get some informations about 
$H_0^{\rm II}$ and $\Omega_0^{\rm II}$ from the time-delay measurement.

In the two single-shell models, by the way, we numerically obtain the 
following ratios ($\alpha_1$ and $\alpha_2$) of ${\cal D}$ and 
$(D_{\rm l})^2$ to the corresponding 
ones in the homogeneous models with $\Omega_0^{\rm I}$ and $H_0^{\rm
I}$, for the representative double quasar: 0957+561 \ ($z_{\rm l} = 0.36, \
z_{\rm s} = 0.41$). 

\noindent For $(\Omega^{\rm I}_0, \Omega^{\rm II}_0, h^{\rm
 I}, h^{\rm II}/h^{\rm I}) = (0.2, 0.56, 0.7, 0.82)$, 
\begin{equation}
  \label{eq:t6}  
\alpha_1 = 1.01, \qquad \alpha_2 = 1.16,
\end{equation}
and $(\Omega^{\rm I}_0, \Omega^{\rm II}_0, h^{\rm
 I}, h^{\rm II}/h^{\rm I}) = (0.2, 0.88, 0.7, 0.82)$,
\begin{equation}
  \label{eq:t7}  
\alpha_1  = 1.14, \qquad \alpha_2 = 1.25.
\end{equation}
If we take $(H_0)_{\rm eff} \simeq 62$ km s$^{-1}$ Mpc$^{-1}$
\citep{falc97} and
adopt $H_0^{\rm I} = 70$ km s$^{-1}$ Mpc$^{-1}$, the ratio (\ref{eq:t4})
is $H_0^{\rm I}/(H_0)_{\rm eff} \simeq 1.13$, which is consistent with
the above values (\ref{eq:t6}) and (\ref{eq:t7}) for $\beta \approx 1$.

\section{Concluding remarks}

In this paper we derived the angular diameter distances from central
and non-central observers in the ``cosmological void models'', for
which we adopted the model parameters necessary to explain
the bulk flow (derived in the previous paper \citep{tom99}), and
showed that the [m, z] relation due to these distances may explain the 
observed deviation of  high-redshift supernovas (SNIa) from the
relation in homogeneous Friedmann models, independently of the 
cosmological constant. This is possible because the void-like low-density
region with a high Hubble parameter gives some ``acceleration'' to
light coming from the high-density region with a low Hubble parameter, 
as if we were in a universe dominated by the positive cosmological constant.

It was found that the remarkable difference between the relation in
the present models and the relation in the
cosmological-constant-dominated model 
appears at epoch $z \simeq 1.0$ or at the earlier stage. The observation
of SNIa around this epoch is, therefore, important to discriminate  
these two models observationally.

The unique property of the present models is that the [m, z] relation
is anisotropic, in contrast to the relation in homogeneous Friedmann
models, and that the systematic image deformation appears for the
sources with $z > z_1^j \ (\sim 0.067)$. The observations about 
the anisotropy and lens effect will also be useful to determine which 
of the two models is better.  

The derivation of distances in models with the intermediate
self-similar region was not treated here, but their behavior is
basicly similar to that in the double-shell models, though their 
analysis is somewhat more complicated.

\appendix
\section{Distances in models with double shells}

\subsection{Distances from the center of the inner region}

The line-elements in the regions V$^{\rm I}$, V$^{\rm II}$ and V$^{\rm
III}$ are given by Eq. (\ref{eq:m1}) with $j = $ I, II and III,
respectively. When S is in V$^{\rm I}$, V$^{\rm II}$ and V$^{\rm
III}$, the angular diameter distance $d_A$ is given by
Eq. (\ref{eq:m2}), Eq. (\ref{eq:m3}) and 
\begin{equation}
  \label{eq:m11}
d_A = a^{\rm III} (\bar{\eta}^{\rm III}_{\rm s}) \sinh \bar{\chi}^{\rm
III}_{\rm s}, 
\end{equation}
respectively, where the negative curvature was assumed also in V$^{\rm
III}$. When S in V$^{\rm I}$ and V$^{\rm II}$, $d_A$ has the same
expressions (for given $\bar{z}^{\rm I}_1$ and $\bar{z}^{\rm II}_{\rm
s}$), as in the single-shell model. When S in V$^{\rm III}$, we have
for the coordinates of the second shell $(\bar{\eta}^j_2,
\bar{\chi}^j_2)$ and those of S, $(\bar{\eta}^{\rm III}_{\rm s}, 
\bar{\chi}^{\rm III}_{\rm s})$ :
\begin{equation}
  \label{eq:m12}
\bar{\eta}^{\rm II}_1 - \bar{\eta}^{\rm II}_2 = \bar{\chi}^{\rm II}_2
- \bar{\chi}^{\rm II}_1,
\end{equation}
\begin{equation}
  \label{eq:m13}
\bar{\eta}^{\rm III}_2 - \bar{\eta}^{\rm III}_{\rm s} =
\bar{\chi}^{\rm III}_{\rm s} - \bar{\chi}^{\rm III}_2,
\end{equation}
where $\bar{\eta}^{\rm II}_2, \ \bar{\eta}^{\rm III}_{\rm s}$ are
related to $\bar{z}^{\rm II}_2, \ \bar{z}^{\rm III}_{\rm s}$ by
\begin{equation}
  \label{eq:m14}
{1 + \bar{z}^{\rm II}_2 \over 1 + \bar{z}^{\rm II}_1} = 
{\cosh \bar{\eta}^{\rm II}_1 -1 \over \cosh \bar{\eta}^{\rm II}_2 - 1},
\end{equation}
\begin{equation}
  \label{eq:m15}
{1 + \bar{z}^{\rm III}_{\rm s} \over 1 + \bar{z}^{\rm III}_2} = 
{\cosh \bar{\eta}^{\rm III}_2 -1 \over \cosh \bar{\eta}^{\rm III}_{\rm 
s} - 1}.
\end{equation}
In the same way as $\bar{z}^{\rm I}_1 = \bar{z}^{\rm II}_1$ in the
previous subsection, we have the equality of the shell redshifts 
$\bar{z}^{\rm II}_2 = \bar{z}^{\rm III}_2$.  Moreover, coordinates
$(\bar{\eta}^{\rm III}_2, \bar{\chi}^{\rm III}_2)$ are connected with 
$(\bar{\eta}^{\rm II}_2, \bar{\chi}^{\rm II}_2)$ by Eqs. (I.A14) and
(I.A15). Accordingly, $D_A$ is uniquely determined for given
$\bar{z}^{\rm I}_1  (= \bar{z}^{\rm II}_1), \ \bar{z}^{\rm II}_2  (=
\bar{z}^{\rm III}_2) \ {\rm and} \ \bar{z}^{\rm III}_{\rm s}$.

Here we calculated the average values of $d_A$ and $d_L \ (= (1 +
z^j)^2)$  defined by Eq. (\ref{eq:r21}) and the $z$ dependence of 
$5 \log d_L$ and $\Delta m$ are shown in Figs. \ref{fig4} and 
\ref{fig6}, respectively.

\subsection{Distances from a non-central observer O in the inner region}

When S in V$^{\rm I}$ and V$^{\rm II}$, $d_A^l, d_A^t$ and
$\varphi_{\rm s}$ are the same as those in the single-shell model.

When S in V$^{\rm III}$, we have
\begin{equation}
  \label{eq:r19a}
d_A^l \equiv d_A(\eta^{\rm III}_{\rm s}, \chi^{\rm III}_{\rm s})
\ \partial \varphi_{\rm s}/ \partial \phi /[\cos (\phi - \varphi_{\rm s})],
\end{equation}
\begin{equation}
  \label{eq:r20a}
d_A^t = d_A(\eta^{\rm III}_{\rm s}, \chi^{\rm III}_{\rm s}) \sin
\varphi_{\rm s}/ \sin \phi,
\end{equation}
where
\begin{equation}
  \label{eq:r21a}
d_A(\eta^{\rm III}_{\rm s}, \chi^{\rm III}_{\rm s}) = a^{\rm
III}(\eta^{\rm III}_{\rm s}) \sinh (\chi^{\rm III}_{\rm s}).
\end{equation}
The relation between $(\eta^{\rm III}_{\rm s}, \chi^{\rm III}_{\rm s},
\varphi_{\rm s})$ and $(\eta^{\rm I}_0, \chi^{\rm I}_0, 0)$ are
 given as follows by specifying $\bar{z}_1^{\rm I}, \bar{z}_2^{\rm
 II}, z^{\rm III}_{\rm s}$ and $\phi$. In the second shell we have
\begin{equation}
  \label{eq:r22}
G(\chi^{\rm II}_2) \equiv \cosh^{-1} \Big({\cosh
\chi^{\rm II}_2 \over h_0^{\rm II}}\Big) - \cosh^{-1} \Big({\cosh 
\chi_1^{\rm II} \over h_0^{\rm II}}\Big) = \eta_1^{\rm II} - \eta_2^{\rm II},
\end{equation}
and in V$^{\rm III}$
\begin{equation}
  \label{eq:r23}
G(\chi^{\rm III}_{\rm s}) \equiv \cosh^{-1} \Big({\cosh
\chi^{\rm III}_{\rm s} \over
h_0^{\rm III}}\Big) - \cosh^{-1} \Big({\cosh \chi_2^{\rm III} \over
h_0^{\rm III}}\Big) 
= \eta_2^{\rm III} - \eta_{\rm s}^{\rm III},
\end{equation}
where
\begin{equation}
  \label{eq:r24}
h_0^{\rm III} = [1 + (\zeta^{\rm III})^2]^{1/2}, \quad \zeta^{\rm III} =
{a_0^{\rm I} \over a_0^{\rm III}} \zeta^{\rm I}.
\end{equation}
The coordinates $(\eta^{\rm III}_2, \chi^{\rm III}_2)$
are related to $(\eta^{\rm II}_2, \chi^{\rm II}_2)$, using
\begin{equation}
  \label{eq:r25}
a_0^{\rm II} y^{\rm II} ({\eta}_2^{\rm II}) \sinh \chi_2^{\rm II} = a_0^{\rm
III} y^{\rm III} ({\eta}_2^{\rm III}) \sinh \chi_2^{\rm III} 
\end{equation}
and 
\begin{equation}
  \label{eq:r26}
a_0^{\rm II} \int_0^{{\eta}_2^{\rm II}} y^{\rm II} (\eta) 
d\eta  = a_0^{\rm III} \int_0^{{\eta}_2^{\rm III}} y^{\rm III} 
(\eta) d\eta.
\end{equation}

Moreover, $\varphi_2$ is derived in V$^{\rm II}$ as
\begin{eqnarray}
  \label{eq:r27}
\varphi_2 = \varphi_1 &+& \tan^{-1} \Big\{{1 \over \zeta^{\rm II}} 
\Big[\sinh^2 \chi_2^{\rm II} + \cosh \chi_2^{\rm II} \sqrt{\sinh^2 \chi_2^{\rm
II} - (\zeta^{\rm II})^2}\Big]\Big\} \cr
&-& \tan^{-1} \Big\{{1 \over 
\zeta^{\rm II}} \Big[\sinh^2 \chi^{\rm II}_1 + \cosh \chi^{\rm II}_1 
\sqrt{\sinh^2 \chi^{\rm II}_1 - (\zeta^{\rm II})^2}\Big]\Big\}
\end{eqnarray}
and in V$^{\rm III}$ we have
\begin{eqnarray}
  \label{eq:r28}
\varphi = \varphi_2 &+& \tan^{-1} \Big\{{1 \over \zeta^{\rm III}} 
\Big[\sinh^2 \chi^{\rm III} + \cosh \chi^{\rm III} \sqrt{\sinh^2 \chi^{\rm
III} - (\zeta^{\rm III})^2}\Big]\Big\} \cr
&-& \tan^{-1} \Big\{{1 \over 
\zeta^{\rm III}} \Big[\sinh^2 \chi^{\rm III}_2 + \cosh \chi^{\rm
III}_2 \sqrt{\sinh^2 \chi^{\rm III}_2 - (\zeta^{\rm III})^2}\Big]\Big\},
\end{eqnarray}
\begin{equation}
  \label{eq:r29}
\varphi_{\rm s} = \varphi (\chi^{\rm III} = \chi^{\rm III}_{\rm s}).
\end{equation}
Thus $\varphi_{\rm s}$ was derived as a function of $\bar{z}_1^{\rm
I}, \ \bar{z}_1^{\rm III},\ z_{\rm s}^{\rm III}$ and $\phi \ (=
\phi_1$ or $\pi -  \phi_1)$.

The author thanks B.P. Schmidt for helpful discussions, and a referee 
for kind comments about past related
works. This work was supported by Grant-in Aid for Scientific Research 
(No. 10640266) from the Ministry of Education, Science, Sports and
Culture, Japan. 


\clearpage

\figcaption[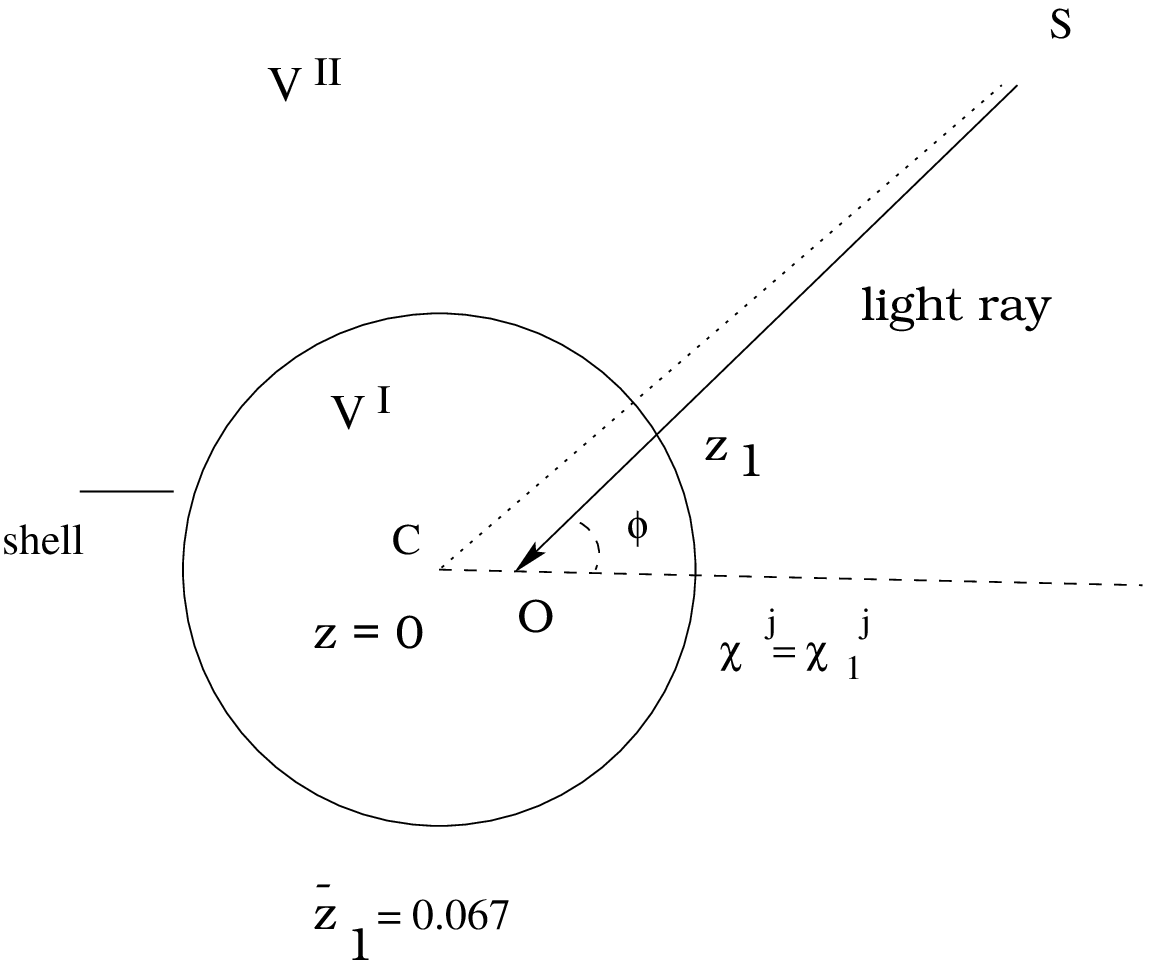]{A model with a single shell. $z$ and $\bar{z}$ are the
 redshifts for observers at O and C.  \label{fig1}}

\figcaption[lens2.ps]{The relation between $d_A$ and $z$. 
A solid line (a) and a short dash line (b) represent the 
 single-shell model with $(\Omega^{\rm I}_0, \Omega^{\rm II}_0, h^{\rm
 I}, h^{\rm II}/h^{\rm I}) = (0.2, 0.56, 0.7, 0.82)$. (a) denotes the
 $\phi$ averaged value by the observer O and (b) denotes the case of
 the virtual observer C. (c) and (d) stand for the homogeneous models
 with $(\Omega_0, \lambda_0) = (0.2, 0)$ and $(0.2, 0.8)$,
 respectively.   \label{fig2}}

\figcaption[lens3.ps]{The relation between $d_A$ and $z$. 
A solid line (a) and a short dash line (b) represent the 
 single-shell model with $(\Omega^{\rm I}_0, \Omega^{\rm II}_0, h^{\rm
 I}, h^{\rm II}/h^{\rm I}) = (0.2, 0.88, 0.7, 0.82)$. The other
 meaning of lines is the same as that in Fig. \ref{fig2}.
 \label{fig3}}

\figcaption[lens4.ps]{The relation between $d_A$ and $z$.  
A long dash line (a) and  a short dash line (b) represent 
the double-shell models with 
 $(\Omega^{\rm I}_0, $ $\Omega^{\rm II}_0, \Omega^{\rm III}_0, $ $h^{\rm
 I}, h^{\rm II}/h^{\rm I}, h^{\rm III}/h^{\rm I})$ $ = (0.2, 0.56, 0.88,
 0.7, 0.92, 0.82),$ $ (0.2, 0.36, 0.56, 0.7, 0.92, 0.82)$,
 respectively, for the observer C. (c) denotes the line by the
 observer C in the single-shell model with $(\Omega^{\rm I}_0, 
\Omega^{\rm II}_0, h^{\rm
 I}, h^{\rm II}/h^{\rm I})$ $ = (0.2, 0.56, 0.7, 0.82)$. (d) denotes the
 Friedmann model with $\Omega_0 = 0.2$.  \label{fig4}}

\figcaption[lens5.ps]{The $z$ dependence (in the single-shell models) of the 
magnitude difference $\Delta m$ relative to the magnitude in the 
Friedmann model with the same $\Omega_0$. Lines (a), (b), (c) and (d)
 stand for the parameters $(\Omega^{\rm I}_0, \Omega^{\rm II}_0, $ $h^{\rm
 I}, h^{\rm II}/h^{\rm I})$ $ = (0.2, 0.56, 0.7, 0.82),$ $ (0.2, 0.88, 0.7,
0.82),$ $ (0.3, 0.56, 0.7, 0.82),$ $ (0.3, 0.88, 0.7, 0.82)$, respectively, 
for $z_1 = 0.067$. (e) denote the case $(0.2, 0.56, 0.7, 0.82)$ for
$z_1 = 0.1$. (f) denotes the homogeneous model with $(\Omega_0,
\lambda_0) = (0.2, 0.8)$.  \label{fig5}}

\figcaption[lens6.ps]{The $z$ dependence (in the double-shell models) of the 
magnitude difference $\Delta m$ relative to the magnitude in the 
Friedmann model with the same $\Omega_0$. Lines (a) and (b) stand for
$(\Omega^{\rm I}_0, \Omega^{\rm II}_0, \Omega^{\rm III}_0, h^{\rm
 I},$ $ h^{\rm II}/h^{\rm I}, h^{\rm III}/h^{\rm I})$ $ = (0.2, 0.36, 0.56,
0.7, 0.92, 0.82),$ $ (0.2, 0.56, 0.88, 0.7, 0.92, 0.82)$,
 respectively, with $z_1 = 0.05$ and $z_2 = 0.1$. (c) is shown for
comparison in the single-shell model with $(\Omega^{\rm I}_0, 
\Omega^{\rm II}_0, h^{\rm I}, h^{\rm II}/h^{\rm I})$ $ = (0.2, 0.56, 
0.7, 0.82)$ and $z_1 = 0.067$.   \label{fig6}}

\figcaption[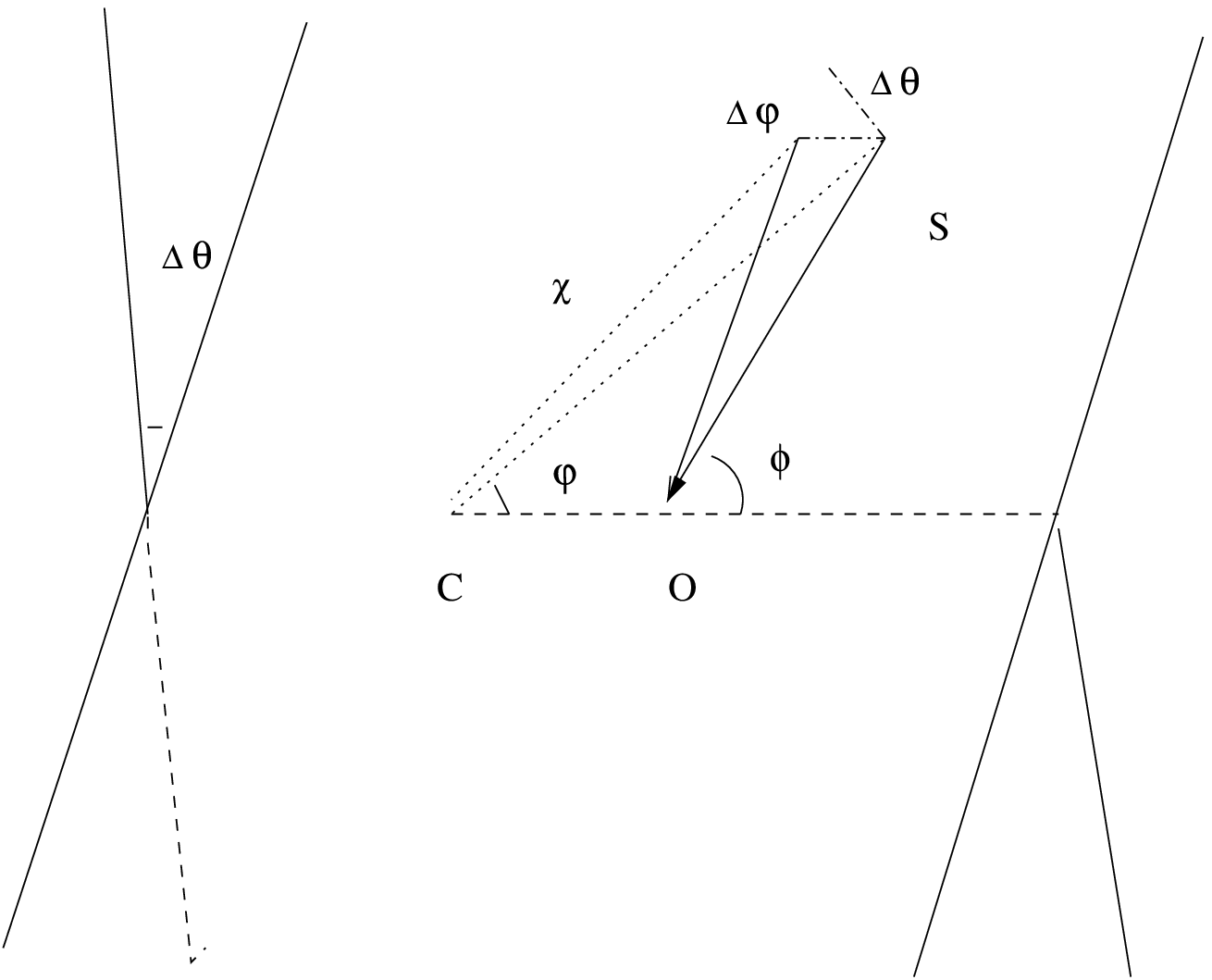]{Planes with constant $\theta$.  \label{fig7}}

\clearpage
\figcaption[lens8.ps]{The $z$ dependence of the
magnitude difference $\Delta m$ relative to the magnitude in the 
Friedmann model with the same $\Omega_0$ for the observer O in the 
single-shell model with $(\Omega^{\rm I}_0, \Omega^{\rm II}_0, h^{\rm
 I}, h^{\rm II}/h^{\rm I}) = (0.2, 0.56, 0.7, 0.82)$. Lines (a), (b) and (c)
 stand for the values averaged for $\phi < \pi/4, \pi/4 < \phi <
3\pi/4,$ and $\phi > 3\pi/4$, respectively. For comparison (d) stands
for the value by the observer C and (e) is for the homogeneous model
with $(\Omega_0, \lambda_0) = (0.2, 0.8)$.  \label{fig8}}

\figcaption[lens9.ps]{The $z$ dependence of the 
magnitude difference $\Delta m$ relative to the magnitude in the 
Friedmann model with the same $\Omega_0$ for the observer O in the 
single-shell model with $(\Omega^{\rm I}_0, \Omega^{\rm II}_0, h^{\rm
 I}, h^{\rm II}/h^{\rm I}) = (0.2, 0.88, 0.7, 0.82)$. The meaning of
suffices are the same as those in Fig. \ref{fig8}.  \label{fig9}}

\figcaption[lens10.ps]{The $z$ dependence of the ellipticity $e$ for the observer
O in the single-shell model with $(\Omega^{\rm I}_0, \Omega^{\rm II}_0, h^{\rm
 I}, h^{\rm II}/h^{\rm I}) = (0.2, 0.56, 0.7, 0.82)$. Lines (b), (c) and (d)
 stand for the values averaged for $\phi < \pi/4, \pi/4 < \phi <
3\pi/4,$ and $\phi > 3\pi/4$, respectively. (a) denotes the value
averaged for $0 < \phi <\pi$.   \label{fig10}}

\figcaption[lens11.ps]{The $z$ dependence of the ellipticity $e$ 
for the observer
O in the single-shell model with $(\Omega^{\rm I}_0, \Omega^{\rm II}_0, h^{\rm
 I}, h^{\rm II}/h^{\rm I}) = (0.2, 0.88, 0.7, 0.82)$. the suffices are 
the same as those in Fig. \ref{fig10}.  \label{fig11}}

\end{document}